\documentclass[11pt]{article}
\usepackage{amsmath, amsfonts, amssymb}
\usepackage{graphics}

\makeatletter
\renewcommand{\@oddhead}{\it \scriptsize Astromony Letters,
 Vol. 32, No. 1, 2006  \hfil} \makeatother
\begin{document}
\newpage
\begin{center}
\huge \rm  OUTER PSEUDORING IN THE GALAXY
\par \vspace{30pt} \large \rm A.M.Mel'nik
\par \small \it anna@sai.msu.ru
\par \rm  http://lnfm1.sai.msu.ru/$^\sim$anna
\par     \it Sternberg Astronomical Institute,
Moscow, Russia \
\vspace{30pt}
\par \large \it to be published in Astronomy Letters, 2006
\par \small \copyright 2006 Pleiades Publishing, Inc.

\end{center}

\renewcommand{\abstractname}{}
\begin {abstract}
The kinematical features of the Sagittarius ($R=5.7$ kpc), Carina
($R=6.5$ kpc), Cygnus ($R=6.8$ kpc), and Perseus ($R=8.2$ kpc) arm
fragments suggest the existence of two spiral patterns rotating
at different angular velocities in the Galaxy. The inner spiral
pattern represented by the Sagittarius arm rotates at the angular
velocity of the bar, $\Omega_{\textrm{b}}=60\pm5$ km s$^{-1}$
kpc$^{-1}$. The outer spiral pattern, which consists of the
Carina, Cygnus and Perseus arms, rotates at a smaller angular
velocity, $\Omega_{\textrm{s}}=12\text{--}22$ km s$^{-1}$
kpc$^{-1}$. The existence of the outer slow tightly wound spiral
pattern and the inner fast spiral pattern can be explained in
terms of the results of numerical simulations of the dynamics of
the outer pseudoring. The OLR of the bar must be located between
the Sagittarius and Carina arms. The Cygnus arm appears as a
connecting link between the fast and slow spiral patterns.

\end{abstract}

\newpage
\section*{\rm \large  INTRODUCTION}

The  kinematics of gas in the central region, infrared photometry,
star counts, and other modern tests prove the presence of the bar
in the Galaxy. Elliptical orbits near the Galactic center allow
many peculiarities of gas kinematics including the co-called 3~kpc
expanding arm to be explained (Peters, 1975; Liszt, Burton,
1980). There is extensive evidence suggesting that the major axis
of the bar is oriented in the direction of
$\theta_{\textrm{b}}=15\text{--}45^\circ$ so that the tip of the
bar nearest to the Sun  lies in the first quadrant. However,
observational estimates of the angular velocity of the bar and
its length are rather uncertain (see Kuijken, 1996; Gerhard, 1996
and references therein).

In this paper we analyze the kinematics and distribution of
OB-associations located within 3 kpc of the Sun. A model of the
Galaxy with an outer pseudoring made up of tightly wound spiral
arms can explain  the unusually small interarm distance
$\lambda=2$ kpc between the Carina and Perseus arm fragments
(Mel'nik et. al., 2001). The kinematics of young stars is clearly
indicative of  slow rotation of the spiral pattern represented by
the Carina, Cygnus, and Perseus arm fragments, which is in a good
agreement with numerical simulations of the dynamics of the outer
pseudoring  (Rautiainen and Salo, 1999, 2000). The galactic model
with fast and slow rotating spiral patterns at  different
Galactocentric distances allows  us to explain the fact that the
kinematics of young stars in the Carina, Cygnus, and Perseus arms
differs from that in the Sagittarius arm.

\section*{\rm \large NUMERICAL SIMULATIONS OF  OUTER PSEUDORINGS
AND CONFRONTING MODELS AND OBSERVATIONS}

Schwarz (1981) was the first to show that  outer rings result
from the evolution of  barred galaxies.  His numerical
simulations  showed  two tightly wound spiral arms making up a
pseudoring to develop in the galactic periphery  after $\sim10$
bar revolutions. The location of the outer pseudoring corresponds
to the outer Lindblad resonance (OLR) of the bar. Schwarz (1981)
found cloud-cloud collisions  to play a crucial role in  the
formation of the outer pseudoring. Shocks in spiral arms emerging
from the bar produce slow flow of gas clouds ($V_R\le1$ km
s$^{-1}$) towards the  OLR and conduce to the gas accumulation in
the OLR region. The  time scale of this process is 10$^9$ yeas.
It determines the time needed for the galaxy to form an outer
pseudoring. Schwarz discovered two types of the outer pseudorings
-- aligned with the bar and perpendicular to it. He associated
them with two types of  periodic orbits near the OLR. Periodic
orbits between the corotation radius (CR) and the OLR are
perpendicular to the bar, whereas orbits outside the OLR are
aligned with the bar.

Buta (1986, 1995) discovered and studied more than 500 galaxies
with outer rings or pseudorings (broken rings), of which 90\%
proved to be barred. Buta and Crocker (1991) found galaxies with
well-defined characteristic features of each pseudoring type,
which can be considered as prototypes. Type $R_1 '$ pseudorings
are made up of spiral arms, which wind $\sim180^\circ$ with
respect to the ends of the bar. These pseudorings are aligned
perpendicular to the bar and always lie inside the OLR of the
bar. Type $R_2 '$ pseudorings are made up of spiral arms, which
wind $\sim270^\circ$ with respect to the ends of the bar. They
are aligned with the bar and lie mostly outside the OLR. Figure~1
shows a schematic view of a typical galaxy with an $R_2 '$
pseudoring from the paper by Buta and Crocker (1991).

The simulation of ring galaxies carried out by Byrd et al. (1994)
shows that first a type $R_1 '$ pseudoring appears  and then it
transforms into a type $R_2 '$ pseudoring. Type $R_2 '$ rings are,
on average, located father away from the galactic center  than
type $R_1 '$ pseudorings, and therefore this sequence of
pseudoring formation further emphasizes the important role that
gas coming from the central region  plays in the formation of
outer rings.

What is the angular rotation velocity, $\Omega_{\textrm{p}}$, of
the tightly wound spiral pattern that forms the  pseudoring? The
above authors implied that  the bar and the outer pseudoring
rotate at the same angular velocity.

Another direction in modeling  barred galaxies led to the
discovery in galactic discs of several patterns rotating at
different angular velocities (Sellwood, 1985; Sellwood and
Sparke, 1988; and other papers). The slowly rotating spiral
pattern was found at some distance from the galactic center.
Tagger et al (1987) and Sygnet et al. (1988) developed a
mechanism of non-linear coupling between the bar and slowly
rotating spiral pattern. They show that the bar and slow spiral
pattern can exchange energy and angular momentum if the CR of the
bar coincides with the inner Lindblad resonance (ILR) of the slow
spiral mode.

Rautiainen and Salo (1999) were the first to find that the outer
pseudoring may comprise spiral arms rotating slower than the bar.
Numerical simulations show that in some cases a galaxy may
possesses both inner and outer spiral arms. Inner arms rotate at
the angular velocity of the bar, whereas  outer tightly wound
arms rotate at a lower velocity. The slowly rotating outer
pseudoring is usually located at the outer edge of the OLR of the
bar. Rautiainen and Salo (1999) also pointed out  possible
non-linear coupling of the two spiral modes so that the OLR of the
inner spiral pattern coincides with the ILR of the outer spiral
pattern. And this mechanism must differ from that suggested by
Tagger (1987). Rautiainen and Salo (2000) performed $N$-body
simulations of the evolution of a barred galaxy  that showed the
outer pseudoring to be usually located between the OLR of the bar
and the CR of the slow spiral mode. Another interesting result of
the above authors was the discovery of  cyclic variations in the
pseudoring morphology. First, a type $R_1 '$ pseudoring forms,
then it acquires the features of a type $R_2 '$ pseudoring, then
again those of a type $R_1 '$ pseudoring, and so on.

\section*{\rm \large UNUSUAL SAGITTARIUS ARM}

In our Galaxy  young  clusters and OB-associations located within
3~kpc of the Sun concentrate toward several regions of intense
star formation, which are traditionally associated with the
Sagittarius--Carina, Cygnus--Orion, and Perseus spiral arms
(Morgan et al., 1952; Becker, 1964; Humphreys, 1979 and other
papers). However, the kinematics of young objects in the Carina,
Cygnus, and Perseus complexes differs from that of young objects
in the Orion and Sagittarius regions ( see Mel'nik et al., 2001
for details).

Fig. 2 shows  regions of intense star formation within 3~kpc of
the Sun and the radial component $V_R$ of the average residual
velocity of young stars in these regions. We computed the
components of the stellar residual velocity along the Galactic
radius-vector $V_R$ and in the azimuthal direction $V_\theta$ as
differences of observed velocities and velocities due to the
circular rotation law and solar motion relative to the centroid
of  OB associations projected onto each direction. We adopted the
parameters of the circular Galactic rotation law and the
components of solar motion from Mel'nik et al. (2001). We set the
solar Galactocentric distance equal $R_0=7.1$ kpc (Dambis et al,
1995; Glushkova et al, 1998).

The radial velocity $V_R$ in the Carina, Cygnus, and Perseus
complexes   is directed toward the Galactic center whereas those
in the Orion and Sagittarius regions $V_R$  are directed away
from the Galactic center. The locations of the Carina, Cygnus, and
Perseus spiral arms correspond to the extreme negative values of
 radial velocities $V_R$, i.e. to the maximum velocities toward
the Galactic center (Fig. 2).

Simulations of the behavior of gaseous clouds in a perturbed
potential performed by Roberts and Hausman (1984) showed that in
trailing spiral arms located inside the CR the radial and
azimuthal velocity components must reach their extreme negative
values at the shock front. This is just the picture  observed in
the Cygnus ($R=6.8$ kpc) and Perseus ($R=8.2$~kpc) arms and it
unambiguously indicates that these spiral arms are trailing and
located inside the CR. We find no statistically significant
variations in the azimuthal residual velocity $V_\theta$  across
the Carina arm ($R=6.5$ kpc). However, the Carina arm is closer
to the Galactic center than the Cygnus and Perseus arms and
therefore it must also lie inside the CR (Fig.~2). Hence all
three fragments of the Carina, Cygnus, and Perseus arms are parts
of a slowly rotating spiral pattern. The requirement that the
Perseus arm is located inside the CR constrains the angular
velocity of this  pattern to $\Omega_{\textrm{p}}<25$~km s$^{-1}$
kpc$^{-1}$ (see Mel'nik, 2003 for  details).

The two regions of intense star formation in Orion and
Sagittarius proved  to be located  in the interarm space. What
triggers star formation in these regions? The Orion region is
probably a spur of the Cygnus arm, because it is located at the
outer edge of the arm. This is consistent with the conclusions of
the  studies of spurs by Weaver (1970) and Elmegreen (1980).

Intense star formation in the Sagittarius region is rather
difficult to explain. This region contains two rich
OB-associations (Sgr OB1 and Ser OB1) and has a size of
$\sim500$~pc in the azimuthal direction. We see no other spiral
arm whose spur could form the complex of young objects in
Sagittarius. Although extinction is extremely strong ($A_V=3^m$)
at heliocentric distances 1\mbox{--}2~kpc in the direction
($l=28^\circ$,  $b=-1^\circ$), it remains rather moderate
($A_V<1^m$)  in the direction ($l=8^\circ$, $b=0^\circ$)  out to
$r=2.5$~kpc and would not prevent the discovery of a "true"
spiral arm (Neckel and Klare, 1980; Efremov and Sitnik, 1988).
There is probably indeed no "true" spiral arm behind the
Sagittarius region.

My principal assumption is that the star-forming region in
Sagittarius  is a fragment  of a trailing spiral arm. It has been
traditionally  interpreted  just in this way, or, to be more
precise, as a constituent part of the Sagittarius-Carina  arm.
However, because of their kinematic differences  the Sagittarius
and Carina arms cannot be considered to be fragments of the same
spiral pattern. The abrupt increase of extinction (from 2 to
4$^m$) at heliocentric distances 2--3~kpc in the field centered
at ($l=18^\circ$, $b=0^\circ$) (Neckel and Klare, 1980) also
suggests that the Sagittarius region is a trailing spiral arm.
Once adopted, this assumption immediately implies that the
Sagittarius arm is a fragment of another spiral pattern, which
rotates at a higher $\Omega_{\textrm{p}}$, in contrast to the
Carina, Cygnus and Perseus arms, which represent a slow spiral
pattern.

Indeed the fact that the radial velocity $V_R$ of young stars in
the Sagittarius arm is directed away from the Galactic center
($V_R=+10$~km s$^{-1}$) indicates that this arm is located
between the CR and OLR of its spiral pattern. Analytical
calculations of stellar motions in  tightly wound trailing spiral
arms show that the direction of the radial  component of velocity
perturbation in the arm depends on the position of the arm with
respect to the  CR: it is directed toward and away from the
Galactic center inside and outside the CR, respectively (Lin et
al., 1969). In this paper we assume all spiral arms to be
trailing.

In a spiral density wave the deviation of the  velocities of
young  objects from circular rotation is due to their coherent
motions along epicycles and the  crowding of the orbits in the
spiral arm corresponds to a certain phase of epicyclic motion. As
long as the density wave exists, the  crowding of the orbits in
the stellar and gaseous subsystems of the disk must correspond to
about the same phase of epicyclic motion and therefore the
allowance for collisions does not alter the conclusion that the
Sagittarius arm is located outside the CR.

The position of the Sagittarius arm between the CR and OLR means
that the CR of this spiral pattern is closer to the Galactic
center than the Sagittarius region ($R_{RC}<5.7$ kpc). Hence the
angular velocity of the rotation of this spiral pattern must be
greater than the angular velocity of Galactic rotation at the
distance of the Sagittarius arm, $\Omega_{\textrm{p}}>38$~km
s$^{-1}$ kpc$^{-1}$.

Thus there are two spiral patterns in the Galaxy, which rotate at
different angular velocities. The spiral pattern that is closer
to the Galactic center rotates faster ($\Omega_{\textrm{p}}>38$
km s$^{-1}$ kpc$^{-1}$) than the spiral pattern located farther
from the Galactic center ($\Omega_{\textrm{p}}<25$~km s$^{-1}$
kpc$^{-1}$).

\section*{\rm \large IDENTIFICATION OF  ARM FRAGMENTS WITH ELEMENTS OF THE
PSEUDORING}

The fact that the Galactic radius-vector drawn at  angles
$\theta=10\text{--}15^\circ$ crosses two spiral arms -- the
Cygnus ($R=6.8$~kpc) and Perseus ($R=8.2$~kpc) fragments --
suggests that the Galaxy possesses a tightly wound and not a
multi-armed spiral pattern (Fig. 2). Fast periodic density
variations along the radius are virtually impossible in the
multi-armed spiral pattern because of the large pitch angle of
spiral arms.

Owing to  the $\sim270^\circ$ winding of the spiral arms with
respect to the bar, type $R_2 '$ pseudorings have regions where
the same galactic radius-vector crosses two tightly wound spiral
arms (Fig. 1). These regions lie at smaller angles $\theta$
relative to the direction of elongation of the bar, i.e., they go
slightly behind the bar (Buta, Crocker, 1991). On the average, the
direction $\theta=10\text{--}15^\circ$ in the Galaxy toward which
the double spiral arms are observed somewhat lags behind the
direction of elongation of the bar,
$\theta_{\textrm{b}}=15\text{--}45^\circ$, and this fact agrees
well with the type $R_2 '$ pseudoring model.

The termination of the Perseus arm in the III quadrant provides
additional evidence for the presence of a type $R_2 '$ pseudoring
in the Galaxy. The $R_2 '$ pseudoring must be stretched along the
bar. Hence the tightly wound spiral arms forming the pseudoring
must terminate in the direction of  bar elongation. In the solar
neighborhood the Perseus arm is the most distant one  from the
Galactic Center. The termination of the Perseus arm at  Per~OB1
association ($\theta=9^\circ$) agrees rather well with the
direction of the bar elongation,
$\theta_{\textrm{b}}=15\text{--}45^\circ$.

Thus the morphology and kinematics of the Galaxy suggest the
presence of a type $R_2 '$ pseudoring. Numerical simulations
indicate that the type $R_2 '$ pseudoring must lie mostly outside
the OLR of the bar. For numerical simulations to be consistent
with the kinematic features of  observed spiral arms, one must
assume that the OLR of the bar  lies  between the Sagittarius and
Carina arms (Fig. 2). The Carina, Cygnus, and Perseus arms, which
represent the slow spiral pattern, would then  be located outside
the OLR, whereas the Sagittarius arm, which represents the fast
spiral pattern, would be located inside the OLR. Such a picture
is consistent with the presence in the Galaxy of  inner and outer
spiral arms. Inner spiral arms must lie between the CR and OLR of
the bar and rotate at the angular velocity of the bar,
$\Omega_{\textrm{b}}$. Outer, tightly wound spiral arms must lie
at the outer edge of the OLR and rotate slower than the bar.
(Rautiainen and Salo, 1999, 2000).

\section*{\rm \large FAST ROTATING SPIRAL PATTERN}

The localization of the OLR of the bar between the fragments of
the Sagittarius ($R=5.7$ kpc) and Carina ($R=6.5$ kpc) arms
allows its angular rotation velocity $\Omega_{\textrm{b}}$ to be
rather accurately determined:

\begin{equation}
\Omega_{\textrm{b}}=\Omega(R_{\textrm{OLR}})+\kappa(R_{\textrm{OLR}})/2
\end{equation}

where $\Omega(R_{OLR})$ is the average angular velocity of disk
rotation and $\kappa(R_{OLR})$ is the epicyclic frequency at the
distance of the OLR. We use the  parameters of the circular law
of Galactic rotation from Mel'nik et al. (2001) to obtain
$\Omega_{\textrm{p}}=60\pm5$~km s$^{-1}$ kpc$^{-1}$. The error in
$\Omega_{\textrm{b}}$ is mostly determined by the accuracy of the
localization of the OLR. In this case no question arises about
the number of spiral arms. Both the bar and associated spiral
pattern must have the mode  m=2.

The CR of the bar rotating at angular velocity of
$\Omega_{\textrm{b}} =60\pm5$ km s$^{-1}$ kpc$^{-1}$ must lie at
the distance of $R_{RC}=3.3\pm0.5$ kpc. Weiner and Sellwood
(1999) obtained similar locations for the CR and OLR of the bar
($R_{\textrm{CR}}=3.6$~kpc and $R_{\textrm{OLR}}\approx R_0$) by
analyzing the ($l\text{--}V$) diagrams of HI and CO distributions
in the central region of the Galaxy.

Fig. 1 clearly shows the difference in the pitch angles of the
spiral arms in the inner and outer galactic regions. The pitch
angle of the spiral arms in the inner and outer regions are close
to 90$^\circ$ and  0$^\circ$, respectively. The change in the
pitch angle of spiral arms near the OLR of the bar was pointed
out by Schwarz (1981), Combes and Gerin (1985), and other
authors. Let us assume that the inner spiral arm begins at the CR
of the bar at (R=3.3~kpc, $\theta=25^\circ$) and passes through
the Sagittarius complex ($R=5.7$ kpc, $\theta=3^\circ$). The
average pitch angle of inner spiral arms must then be equal to
$i=55^\circ$.

\section*{\rm \large SLOW ROTATING SPIRAL PATTERN}

Let us now consider the slowly rotating spiral pattern, which is
represented by the Carina, Cygnus, and Perseus arm fragments. In
most of the cases pseudorings in galaxies  consist of two tightly
wound spiral arms (Buta, 1995). In the Galaxy such arms are
probably represented by the Carina and Perseus arm fragments. For
the two-armed model ($m_s=2$) the pitch angle of the outer spiral
arms must be equal to $i=5^\circ$ (Mel'nik et al., 2001).

The Carina arm ($R=6.5$ kpc) is likely to be located near the ILR
of the slowly rotating  spiral pattern. This is evidenced by the
specific features of its kinematics and the lack of other arms of
the slow spiral pattern at  smaller $R$. The fast spiral pattern
is already present at  $R=5.7$ kpc and therefore the slow spiral
pattern may extend inward only down to Galactocentric distances
$R\approx 5.7$ kpc (Mel'nik et al., 2005).

The kinematic peculiarity of the Carina arm -- namely the absence
of significant cross-arm variations of   the azimuthal velocity
component -- probably suggests that the arm has  degenerated into
a ring. It is very difficult to explain the presence of cross-arm
gradient of radial velocity combined with the absence of
cross-arm gradient of azimuthal velocity. In a spiral density wave
both $V_R$ and $V_\theta$ velocity components  must necessarily
exhibit cross-arm variations. Both velocity gradients are due to
the same cause: the change of the direction of  elongation of
orbits in the disk plane with  increasing Galactocentric distance
$R$.

The observed gradient of the radial velocity $V_R$ across the
Carina arm may be due to  the contraction of orbits located near
the ILR. In the ILR region gaseous clouds must effectively lose
angular momentum and acquire radial velocity directed toward the
Galactic center, while  organized motions in the azimuthal
direction may be absent.

Let us assume that the ILR of the slow spiral mode is located at
the Galactocentric distance of the Carina arm $R=6.5$ kpc. If we
knew the number of spiral arms $m_s$ in the outer spiral pattern,
we could immediately compute the corresponding pattern speed
$\Omega_{\textrm{s}}$:

\begin{equation}
\Omega_{\textrm{s}}
=\Omega(R_{\textrm{ILR}})-\kappa(R_{\textrm{ILR}})/m_s
\end{equation}

where $\Omega(R_{\textrm{ILR}})$ and $\kappa(R_{\textrm{ILR}})$
are  the average angular velocity of disk rotation and the
epicyclic frequency, respectively, calculated at the distance of
the ILR.

In the Galaxy we are dealing with a small number of tightly wound
spiral arms. However, we cannot be sure that the Galaxy has 2 and
not 4 arms. The $m_s=2$ and $m_s=4$ models imply the  pattern
speeds of $\Omega_{\textrm{s}}=12\pm2$  and
$\Omega_{\textrm{s}}=22\pm2$ km s$^{-1}$ kpc$^{-1}$,
respectively. Both values obey the inequality
$\Omega_{\textrm{s}}<25$ km s$^{-1}$ kpc$^{-1}$, which expresses
the requirement that the Perseus arm is located inside the CR.
Note that in the two-armed model ($m_s=2$) the CR of the slow
spiral mode is located at the Galactocentric distance of
$R_{\textrm{RC}}\approx15\text{--}17$ kpc, whereas in the
four-armed model ($m_s=4$) it is located at
$R_{\textrm{RC}}\approx10$ kpc. It is rather difficult  to choose
one of the two models, but in any case the angular velocity of
rotation of the slow spiral pattern must lie in the interval of
$\Omega_{\textrm{s}}=12\text{--}22$ km s$^{-1}$ kpc$^{-1}$.

Fig. 3 shows a schematic view of the  spiral structure of the
Galaxy under the assumption of two-armed symmetry. We set the
angle between  the  major axis of the bar and the solar direction
equal to $\theta_{\textrm{b}}=25^\circ$ and the  axial ratio of
the bar to be equal to 3:1 (Gerhard, 1996; Weiner, Sellwood,
1999). The pith angles of the inner and outer spirals are equal to
$i=55^\circ$ and $i=5^\circ$, respectively. The dashed line shows
the position of the OLR of the bar. The locations of the star-gas
complexes within 3 kpc of the Sun are carried over from Fig. 2.

\section*{\rm \large THE CYGNUS ARM AS A CONNECTING LINK BETWEEN
THE FAST AND SLOW SPIRAL PATTERNS}

Thus the Galaxy possesses a fast ($\Omega_{\textrm{p}}=60\pm5$ km
s$^{-1}$ kpc$^{-1}$) and a slow
($\Omega_{\textrm{s}}=12\text{--}22$ km s$^{-1}$ kpc$^{-1}$)
spiral patterns. How to  explain such a good coupling of the ends
of these two patterns? Complexes of young objects in Sagittarius
($\theta=3^\circ$) and Carina ($\theta=-15^\circ$) are
$\Delta\theta=18^\circ$, or only 5\% of the circumference, apart.
It is this accurate coupling that gave rise to the emergence of
the notion of the Sagittarius--Carina arm, although different
parts of this arm  move at an angular velocity of
$\Delta\Omega=40\text{--}50$ km s$^{-1}$ kpc$^{-1}$ relative to
each other.

We can answer this question if we assume that both spiral
patterns are periodically deformed and adjust to each other. The
Cygnus arm is very much like a connecting link between the fast
and slow spiral patterns. In Fig. 3  the Cygnus arm and the arm
fragment that is symmetric to it are located at the places where
the inner and outer spirals join each other forming specific
"cap-peaks". These "cap-peaks" make outer spirals appear to lead
the inner spirals in  Galactocentric angle. Sellwood and Spark
(1988) found similar connecting links  between the fast rotating
bar and slow rotating spiral pattern.

It is interesting that the ILR of the slow spiral pattern
($R\approx6.5$ kpc) is located near the OLR of the fast spiral
pattern ($R\approx6.1$ kpc). Rautiainen and Salo (1999, 2000)
pointed out the close location of the resonances, the OLR and ILR,
of the two spiral patterns in numerical simulations. Gaseous
clouds, which may accumulate in the narrow ring between the OLR
of the inner and the ILR of the outer spiral mode must play
important role in the mechanism of coupling of  resonances
OLR\mbox{--}ILR. The fast spiral mode forms a flow of gaseous
clouds from the center toward the periphery (Schwarz, 1981),
whereas the slow spiral mode produces a flow  in the opposite
direction, from the periphery toward the center. Possibly two
spiral modes periodically exchange gaseous clouds and amplify
each other.

\section*{\rm \large CONCLUSIONS}

The kinematic features of  the Sagittarius, Carina, Cygnus, and
Perseus arms suggest that the Galaxy possesses  two spiral
patterns rotating at different angular velocities. The
Sagittarius arm is a fragment of a spiral pattern that rotates at
a higher angular velocity than the  pattern represented by the
Carina, Cygnus, and Perseus arms. The presence of a slow outer
tightly wound spiral pattern and a fast inner spiral pattern can
be explained using   numerical simulations of the dynamics of
outer pseudorings  (Rautiainen and Salo, 1999, 2000). The OLR of
the bar must then be located between the Sagittarius and Carina
arms. In this case the Carina, Cygnus, and Perseus arms, which
represent the slow spiral pattern, must be located outside the
OLR, whereas  the Sagittarius arm, which  represents the fast
spiral pattern, must lie  inside the OLR of the bar.

The localization of the OLR of the bar between the Sagittarius
and Carina arms allows us to determine the angular velocity of
bar rotation $\Omega_{\textrm{b}} =60\pm5$ km s$^{-1}$
kpc$^{-1}$, and the radius of bar coratation
$R_{\textrm{RC}}=3.3\pm0.5$ kpc. The angular velocity of the
rotation of the slow spiral pattern must lie in the interval
$\Omega_{\textrm{s}}=12\text{--}22$ km s$^{-1}$ kpc$^{-1}$.

\section*{\rm \large ACKNOWLEDGMENTS}

I am grateful to A. V. Zasov, I. I. Pasha, Yu. N. Efremov, and A.
S. Rastorguev for interesting discussion and useful remarks. This
work was supported by the Russian Foundation for Basic Research
(projects nos.~~03\mbox{-}02\mbox{-}16288 and
04\mbox{-}02\mbox{-}16689), the Council for the Program of
Support for Leading Scientific Schools (project no.
NSh.389.2003.2), and the State Science and Technology Astronomy
Program.

\section*{\rm \large REFERENCES}

\begin{enumerate}

\item W. Becker, { \it The Galaxy and the Magellanic Clouds, IAU Symp.20}, Ed. F.
J. Kerr (Canberra: Austral. Acad. Sci., 1964), p.16

\item R. Buta, Astrophys. J. Suppl. Ser. {\bf 61}, 609 (1986).

\item R. Buta, Astrophys. J. Suppl. Ser. {\bf 96}, 39 (1995).

\item R. Buta and D. A. Crocker, Astron. J. {\bf 102}, 1715 (1991).

\item G. Byrd, P. Rautiainen, H. Salo, R. Buta and D. A. Crocker ,
Astron. J. {\bf 108}, 476 (1994).

\item F. Combes and M. Gerin, Astron. Astrophys. {\bf 150}, 327 (1985).

\item A. K. Dambis, A. M. Mel'nik, and A. S. Rastorguev, Astron. Lett.
{\bf 21}, 291 (1995).

\item Yu.N. Efremov and T.G. Sitnik, Astron. Lett. {\bf 14}, 347 (1988).

\item D. M. Elmegreen, Astrophys. J. {\bf 242}, 528 (1980).

\item O. E. Gerhard, { \it Unsolved Problems of the Milky Way, IAU Symp. 169},
Ed. L. Blitz, P. Teuben (Dordrecht: Kluwer, 1996), p.79.

\item E. V. Glushkova, A. K. Dambis, A. M. Mel'nik, and A. S.
Rastorguev, Astron. Astrophys. {\bf 329}, 514 (1998).

\item R. M. Humphreys, { \it The Large Scale Characteristics of the Galaxy,
IAU Symp. 84}, Ed. W. B. Burton (Dordrecht: D. Reidel Publ. Co.,
1979), p.93.

\item K. Kuijken,  ASP Conference Ser. 91, 504, (1996).

\item C. C. Lin, C. Yuan, and F. H. Shu, Astrophys. J. {\bf 155}, 721 (1969).

\item H. S. Liszt and W. B. Burton, Astrophys. J. {\bf 236}, 779 (1980).

\item A. M. Mel'nik, Astron. Lett. {\bf 29}, 304 (2003).

\item A. M. Mel'nik, Astron. Lett. {\bf 31}, 80 (2005).

\item A. M. Mel'nik, A. K. Dambis, and A. S. Rastorguev, Astron. Lett.
{\bf 27}, 521 (2001).

\item W. W. Morgan, S. Sharpless, and D. Osterbrock, Astron. J. {\bf 57}, 3
(1952).

\item Th. Neckel and G. Klare, Astron. Astrophys. Suppl. Ser. {\bf 42}, 251
(1980).

\item W. L. Peters, Astrophys. J. {\bf 195}, 617 (1975).

\item P. Rautiainen and H. Salo, Astron. Astrophys. {\bf 348}, 737 (1999).

\item P. Rautiainen and H. Salo, Astron. Astrophys. {\bf 362}, 465 (2000).

\item W. W. Roberts, and M. A. Hausman, Astrophys. J. {\bf 277}, 744 (1984).

\item M. P. Schwarz, Astrophys. J. {\bf 247}, 77 (1981).

\item J. A. Sellwood, Mon. Not. R. Astron. Soc. {\bf 217}, 127 (1985).

\item J. A. Sellwood and L. S. Sparke, Mon. Not. R. Astron. Soc. {\bf 231},
25 (1988).

\item J. F. Sygnet, M. Tagger, E. Athanassoula, R. Pellat, Mon. Not. R.
Astron. Soc. {\bf 232}, 733 (1988).

\item M. Tagger, J. F. Sygnet, E. Athanassoula, R. Pellat, Astrophys.
J. (Letters) {\bf 318}, L43 (1987).

\item H. Weaver, { \it Interstellar Gas Dynamics, IAU Symp. 39}, Ed. H. Habing
(Dordrecht: D. Reidel Publ. Co., 1970), p.22 .

\item B. J. Weiner and J. A. Sellwood, Astrophys. J. {\bf 524}, 112 (1999).
\end{enumerate}

\newpage
\begin{figure}[t]
\includegraphics{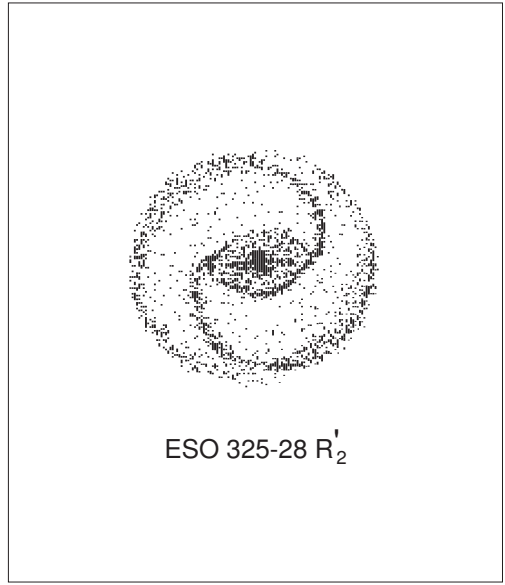}  \vspace{8.0cm} \caption{Image of a typical galaxy
with pseudoring $R_2 '$ from  Buta and Crocker
(1991) (a mirror copy).\hfill}
\end{figure}
\newpage
\begin{figure}[t]
\includegraphics{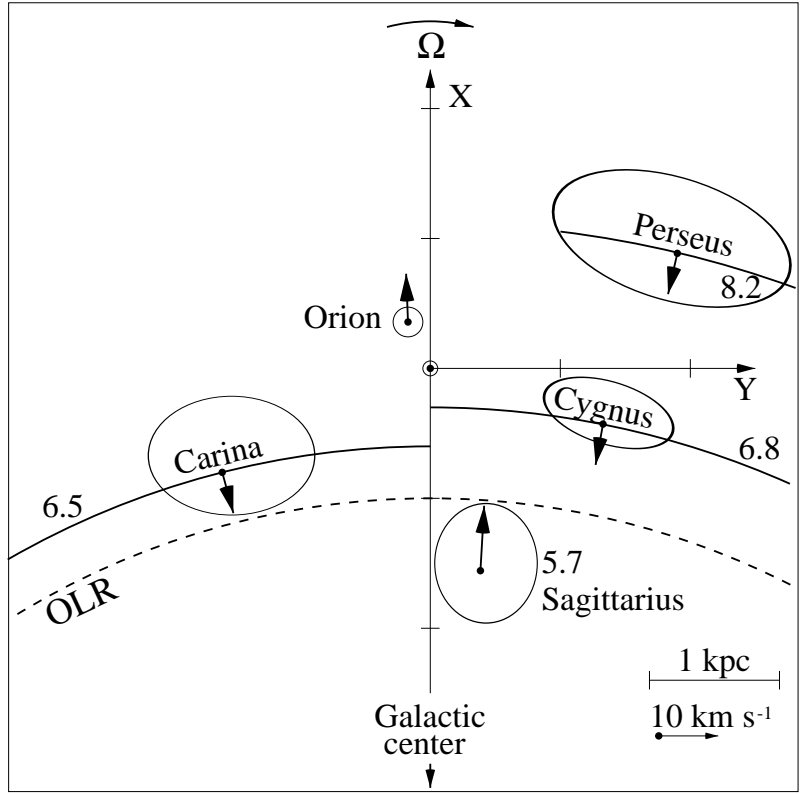}  \vspace{10.0cm} \caption{Regions of intense star
formation in the Galaxy and their average radial velocity $V_R$.
The locations of the Carina, Cygnus, and Perseus spiral arms
correspond to  the extreme negative values of  radial velocity
$V_R$. The dashed line shows the location of the OLR of the bar.
The numbers indicate Galactocentric distances in kpc. \hfill}
\end{figure}
\begin{figure}[t] \includegraphics{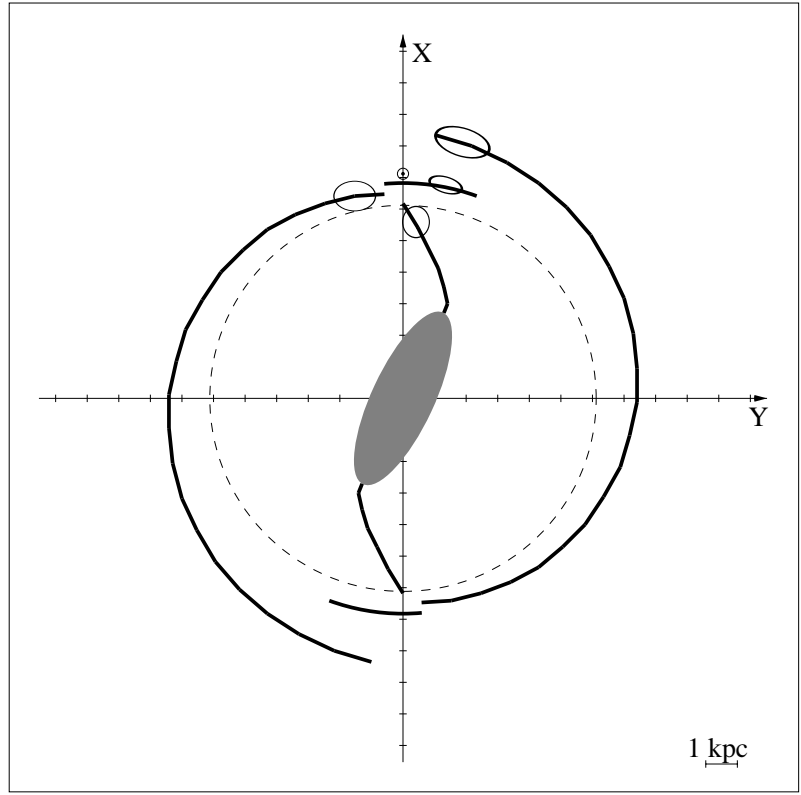}  \vspace{11.0cm}
\caption{Schematic view of the spiral structure of the Galaxy.
The pitch angles of the inner and outer spiral arms are equal to
$i=55^\circ$ and $i=5^\circ$, respectively. The position of the
Sun is indicated by the traditional sign. The positions of
star-gas complexes within 3 kpc are shown by ovals as  in Fig. 2.
The dashed line shows the location of the OLR of the bar.\hfill}
\end{figure}

\end {document}